\documentclass{sig-alternate}

\newcommand{\R}{\mathbb{R}}

\newcommand{\Fc}{\mathcal{F}}

\newcommand{\Uc}{\mathcal{U}}

\newcommand{\Zc}{\mathcal{Z}}

\newcommand{\T}{\top}

\newcommand{\uhatMAP}{\widehat {u}_{\text{MAP}}}

\newcommand{\ie}{i.e.~}

\newdef{assumption}{Assumption}
\newdef{proposition}{Proposition}
\newdef{definition}{Definition}
\newdef{problem}{Problem}

\begin{document}
%
\conferenceinfo{HiCoNS}{April 15-17, 2014, Berlin, Germany}
\title{Fundamental Limits of Nonintrusive Load Monitoring\titlenote{The work presented is supported by the NSF
Graduate Research Fellowship under grant DGE 1106400, NSF
CPS:Large:ActionWebs award number 0931843, TRUST (Team for Research in
Ubiquitous Secure Technology) which receives support from NSF (award
number CCF-0424422), and FORCES (Foundations Of Resilient
CybEr-physical Systems), the European Research Council
   under the advanced grant LEARN, contract 267381, a postdoctoral grant from the Sweden-America
   Foundation, donated by ASEA's Fellowship Fund, and  by a postdoctoral
   grant from the Swedish Research Council.
}}

%
%
%
%
%

\numberofauthors{4} 
%
\author{
%
%
\alignauthor
Roy Dong\\
       \affaddr{Dept. of Electrical Engineering and Computer Sciences}\\
       \affaddr{UC Berkeley}\\
       \affaddr{Berkeley, CA, USA}\\
       \email{\normalsize{roydong@eecs.berkeley.edu}}
\alignauthor
Lillian Ratliff\\
       \affaddr{Dept. of Electrical Engineering and Computer Sciences}\\
       \affaddr{UC Berkeley}\\
       \affaddr{Berkeley, CA, USA}\\
       \email{\normalsize{ratliffl@eecs.berkeley.edu}}
\and  
\alignauthor
Henrik Ohlsson\\
       \affaddr{Dept. of Electrical Engineering and Computer Sciences}\\
       \affaddr{UC Berkeley}\\
       \affaddr{Berkeley, CA, USA}\\
       \email{\normalsize{ohlsson@eecs.berkeley.edu}}
\alignauthor
S. Shankar Sastry\\
       \affaddr{Dept. of Electrical Engineering and Computer Sciences}\\
       \affaddr{UC Berkeley}\\
       \affaddr{Berkeley, CA, USA}\\
       \email{\normalsize{sastry@eecs.berkeley.edu}}
}

\maketitle
\begin{abstract}
Provided an arbitrary nonintrusive load monitoring (NILM) algorithm, we seek bounds on the probability of distinguishing between scenarios, given an aggregate power consumption signal. We introduce a framework for studying a general NILM algorithm, and analyze the theory in the general case. Then, we specialize to the case where the error is Gaussian. In both cases, we are able to derive upper bounds on the probability of distinguishing scenarios. Finally, we apply the results to real data to derive bounds on the probability of distinguishing between scenarios as a function of the measurement noise, the sampling rate, and the device usage.
\end{abstract}



\terms{Theory}
\keywords{nonintrusive load monitoring (NILM); energy disaggregation; performance bounds}

\vfill\eject

\section{Introduction}
\label{sec:introduction}
Nonintrusive load monitoring (NILM) is a general term which refers to determining the energy consumption of individual devices, or statistics of the energy consumption signal, without installing individual sensors at the plug level. 
The goals of different NILM algorithms include event detection, \ie determine when certain devices switch states, and energy disaggregation, \ie  recovering the power consumption signals of each device in its entirety from the aggregate signal. In many cases, we would like to have the latter for many households, but installing sensors on every plug in each house is prohibitively expensive and intrusive. 
For example, studies have shown that merely providing users feedback on their energy consumption patterns is sufficient to improve their consumption behaviors \cite{Gardner2008,Laitner2009,Armel2013}. Forecasts predict that 20\% savings in residential buildings are attainable with the use of personalized recommendations based on disaggregated data. Additionally, these savings are sustainable over long time periods, and are not transient effects of introducing new interfaces to users. These device-level measurements can further be used for strategic marketing of energy-saving programs and rebates, both improving efficacy of the programs and reducing costs.


NILM algorithms can 
 help guide regulation for privacy policies in advanced metering infrastructures (AMIs) \cite{Cardenas2012}. Analyzing NILM algorithms is a way to determine how much device-level information is contained in an aggregate signal. This information is critical to understanding the privacy concerns in AMIs and which parties should have access to aggregate power consumption data.  
Further, NILM algorithms can provide a good benchmark for defining privacy risk; the state--of--the--art NILM algorithm may be a reasonably conservative model for an adversary. For example, if we use the framework defined in \cite{Dong2013,Dong2013a}, we can analyze how much energy disaggregation an adversary can achieve with a prior on the device usage patterns and models for individual devices.

Technologies and algorithms are constantly evolving, and to the best of our knowledge, there has not yet been an attempt to analyze the fundamental limits of NILM algorithms. 
An understanding of the fundamental limits can provide a theoretical guarantee of privacy, if we conclude that disaggregation is impossible in a certain scenario. It can be used in the design of AMIs, by determining a minimum sampling rate, sensor accuracy, and network capacity to achieve a desired goal. Further, it may allow us to determine how many measurements actually need to be stored and transmitted.


In this paper, we study the fundamental limits of NILM algorithms. We consider a building containing a number of devices. Given the aggregate power consumption of these devices, we would like to distinguish between two scenarios, e.g. whether or not a light turns on, or whether it was a toaster or kettle that turned on. In particular, provided an arbitrary NILM algorithm, we seek bounds on the probability of distinguishing two scenarios given an aggregate power consumption signal. Additionally, once we have this theory developed for two scenarios, we generalize to find an upper bound on the probability of distinguishing between a finite number of scenarios. 
With this theory of the fundamental limits of NILM in hand, we address questions about the possibility of NILM in the context of AMIs. Further, using high-frequency, high-resolution measurements of power consumption signals of common household devices as the ground truth, we analyze the probability of successfully identifying common scenarios in a household. We also analyze the tradeoff between successful NILM and sensor/model accuracy, as well as sampling rate. 

This paper is organized as follows. In Section~\ref{sec:background}, we review the relevant literature. We formulate the NILM problem  and the model of NILM algorithms in Sections~\ref{sec:problem} and \ref{sec:framework} respectively. In Section~\ref{sec:theory}, we discuss the fundamental limits of NILM algorithms. We derive bounds on distinguishing a finite number of scenarios using a classical hypothesis testing framework. In Section~\ref{sec:applications}, we focus on the case where the NILM model is deterministic with additive Gaussian noise. We derive analytical expressions for bounds on the probability of distinguishing scenarios. We apply the theory to real--data gathered on a number of household appliances in Section~\ref{sec:real_data}. Finally, in Section~\ref{sec:conclusion}, we make concluding remarks.




\section{Background}
\label{sec:background}

The problem of NILM is essentially a single-channel source separation problem: determine the power consumption of individual devices given their aggregated power consumption. 
The source separation problem has a long history in information theory and signal processing and well known methods include the infomax principle \cite{Bell1995}, which tries to maximize some output entropy, the maximum likelihood principle \cite{Cardoso1997}, which uses a contrast function on some distribution on the source signals, and a time-coherence principle \cite{Belouchrani1997}, which assumes time-coherence of the underlying source signals. These often lead to formulations which use some variation of a principle components analysis (PCA) or independent component analysis (ICA).

The most common applications of the source separation theory is to audio signals and biomedical signals.  For these applications, it is often assumed that source signals are i.i.d.  stationary processes.  We note that power consumption signals are very different from these types of signals. The power consumption of a device has strong temporal correlations and are not stationary, e.g.~whether or not a device is on at a given time is correlated with whether or not it was on an instant ago, and the mean power consumption signal changes with the state of the device. 
The algorithmic and theoretical development in source separation have therefore not been successfully applied to NILM and most methods for NILM are rather different to those developed for classical source separation.

The field of NILM is much younger than source separation and most  development has focused on algorithms. We briefly outline a few approaches here. 
One approach has focused on the design of hardware to best detect the signatures of distinct devices \cite{Leeb1995,Gupta2010,Froehlich2011}, but algorithms to handle the hardware's measurements are still an open problem. Another approach which has been taken by much of the machine learning community is to use hidden Markov models (HMMs), or some variation, to model individual devices \cite{Kolter2011,Kolter2012,Parson2012}; energy disaggregation can be done with an expectation maximization (EM) algorithm. In recent publications \cite{Dong2013,Dong2013a}, we model individual devices as dynamical systems and use adaptive filtering. These are a few examples of concrete algorithms for NILM. For a more comprehensive review, we refer the reader to \cite{Armel2013}.

The discussion  presented in this paper focus on the theoretical limitations of an arbitrary NILM algorithm. To the best of our knowledge, there has not been any previous work attempting to model the NILM problem in its full generality and derive theoretical bounds. The work is inspired  by recent work in differential privacy \cite{Dwork2006, Ny2012}. The underlying goal of differential privacy is to model privacy in a fashion that encapsulates arbitrary prior information on the part of the adversary and an arbitrary definition of what constitutes a privacy breach. 
The theory of differential privacy can be extended to give similar, but weaker, bounds to those derived in this paper.

\section{The problem of NILM}
\label{sec:problem}
As mentioned in Section \ref{sec:introduction}, NILM has a variety of end uses. For each of these potential applications, the statistics of interest may be different. Thus, when we state the problem of NILM, we remain as general as possible to accommodate all these applications.

We are given an aggregate power consumption signal for a building. Let $y[t] \in \R$ denote the value of the aggregate power consumption signal at time $t$ for $t = 1,\dots,T$, and let $y \in \R^{T}$ refer to the entire signal. This signal is the aggregate of the power consumption signal of several individual devices:
\begin{equation}
	y[t] = \sum_{i = 1}^D y_i[t] \text{ for } t = 1,\dots,T
\end{equation}
where $D$ is the number of devices in the building and $y_i[t]$ is the power consumption of device $i$ at time $t$.

There are many possible goals of NILM. For example, the energy disaggregation problem is to recover $y_i$ for $i = 1,2,\dots,D$ from $y$. Another goal commonly studied is to recover information about the $y_i$ from $y$, such as when lights turn on or the power consumption of the fridge over a week.

Generally, we will refer to the phenomena we wish to distinguish as scenarios throughout this paper.

\section{Model of NILM algorithms}
\label{sec:framework}
In this section, we outline a general framework for analyzing the problem outlined in Section \ref{sec:problem}. 
At a high level, the framework can be summarized as follows. First, any NILM method must choose some representation for individual devices; these can be seen as functions from some input space to $\R^{T}$. Depending on the purpose of the NILM algorithm, the input space will vary; essentially, scenarios we wish to distinguish should correspond to different inputs in the input space. Then, we describe NILM algorithms as functions on the observed aggregate signal. The definition is meant to be general and hold across both generative and discriminative techniques. 

\subsection{Aggregate device model}

Formally, let $(\Omega, \Fc, P)$ denote our probability space. As in Section \ref{sec:problem}, $D$ denotes the number of devices and $T$ denotes the length of our observed power signal. 

Let $\Uc_i$ denote the input space for for the $i$th device. Inputs represent scenarios we wish to distinguish. The output space, representing the power consumption signal of an individual device, is $\R^T$ for every device. Then, the model associated with the $i$th device can be denoted as $G_i : \Uc_i \times \Omega \rightarrow \R^T$. Here, we have the condition that, for any $u_i \in \Uc_i$, $G_i(u_i,\cdot)$ is a random variable. Finally, let $\Uc = \Uc_1 \times \Uc_2 \times \dots\ \times \Uc_D$, and let $G : \Uc \times \Omega \rightarrow \R^T$ be defined as $G( (u_1,u_2,\dots,u_D), \omega ) = \sum_{i = 1}^D G_i( u_i, \omega )$. Here, $G$ denotes our aggregated system, \ie  the model of our building.

\begin{assumption}
Given that the input is $u \in \Uc$, the distribution of the power consumption is $G(u,\cdot)$.
\end{assumption}

We emphasize the generality of this framework. Many state-of-the-art methods can be formulated in this framework. For example, factorial hidden Markov model methods \cite{Kolter2011,Kolter2012,Parson2012} can be thought of as single-input, single-output systems where the input is the state of the underlying Markov chains. The Markov transition probabilities become a prior on the input signal. In previous work \cite{Dong2013,Dong2013a}, we formulated the models as dynamical systems whose inputs are real-valued and correspond to the device usage. 
Thus, we now have a general way of expressing different models of devices in a NILM problem.

\subsection{NILM algorithms}
An algorithm for NILM will be a function of our observed aggregate power consumption signal. Its result will depend on the goal of the algorithm, and the end use of the algorithm output. For example, it could be the set of possible estimated disaggregated energy signals, $\{\widehat {y_i}\}_{i = 1}^D$, or the set of possible discrete event-labels on our time-series data, or a set of statistics on the disaggregated data.

More formally, let $S$ represent some NILM algorithm and $\Zc$ represent its output space, discussed above. Then, the algorithm could be thought of as a function $S : \R^T \rightarrow \Zc$. We will analyze a general $S$ in the following section.

\section{Fundamental limits of NILM}
\label{sec:theory}
In this section, we derive an upper bound on the probability of successfully distinguishing two scenarios with any NILM algorithm. Then, we extend these results to handle the case where we wish to upper bound the probabilities of distinguishing a finite set of scenarios, as well as two collections of scenarios. 
Note that in our framework, scenarios correspond to inputs to our device models, and we will use the two terms interchangeably.

\vfill\eject

\subsection{Distinguishing two scenarios}

First, fix any two inputs $v_0, v_1 \in \Uc$ which we wish to distinguish. For example, we may pick $v_0$ and $v_1$ so that they differ only in the usage of one device. In that case, we are analyzing the difference in observed output caused by whether or not, say, a microwave turns on in the morning. Alternatively, we may choose inputs that correspond to more dissimilar scenarios, such as whether or not a household uses an air conditioner at all. The choice of $v_0, v_1$ depends on which scenarios we wish to distinguish in our NILM algorithm. 

As mentioned previously, let $S : \R^T \rightarrow \Zc$ denote any NILM algorithm. Then, let $I : \Zc \rightarrow \{0,1\}$ be an indicator for whether or not an algorithm output satisfies some condition. For example, $I$ could output 1 if a particular discrete phenomena, e.g.~a light turning on, is detected in the algorithm output, and 0 otherwise. Or, $I$ could output 1 if the estimated power consumption signals of individual devices lies in a certain set. 

Suppose that this indicator is supposed to capture whether our algorithm believes the input is $v_0$ or $v_1$. That is, $(I \circ S)$ should output 1 if the NILM algorithm believes the input is $v_1$ and 0 if it believes the input is $v_0$. For this reason, from this point forward we will refer to $I$ as our discriminator.

\begin{assumption}
	$(I \circ S)$ is measurable, \ie  $(I \circ S)^{-1}(\{1\})$ is a measurable set in $\R^T$, with respect to the Borel field on $\R^T$.
\end{assumption}

We note that this is a reasonable assumption, as most, if not all, NILM algorithms in practice will be a finite composition of measurable functions.

Additionally, we note that this is a very conservative understanding of an NILM algorithm. In general, these algorithms are not be designed simply to distinguish between $v_0$ and $v_1$, and are likely not to be optimal in this regard. 
Thus, by analyzing an optimal $(I \circ S)$, we have a conservative upper bound on the probability of distinguishing $v_0$ and $v_1$.

Thus, we can formulate this in classical hypothesis testing frameworks seen in the statistics literature \cite{Papoulis1991}.

Let $y$ denote our observed signal. Suppose that $G(v_0,\cdot)$ has a probability density function (pdf) $f_0$ and similarly $G(v_1,\cdot)$ with $f_1$. Let our likelihood ratio be defined as:
\begin{equation}
L(y) = \frac{f_1(y)}{f_0(y)}
\end{equation}

The maximum likelihood estimator (MLE) finds the input that maximizes the likelihood of our observations.
The MLE is given by:
\begin{equation}
	\widehat {u}_{\text{MLE}}(y) =
	\begin{cases}
		v_1 \text{ if } L(y) \geq 1 		\\
		v_0 \text{ otherwise}
	\end{cases}
\end{equation}

If we have a prior $p$ on the probability of $v_0$ or $v_1$ as inputs, we can find the maximum a posteriori (MAP) estimate. 
This finds the input that is most likely given our observations and prior.  
The MAP is:
\begin{equation}
	\widehat {u}_{\text{MAP}}(y) =
	\begin{cases}
		v_1 \text{ if } L(y) \geq \frac{p(v_0)}{p(v_1)} 		\\
		v_0 \text{ otherwise}
	\end{cases}
\end{equation}
Note that this prior can be a discrete distribution or a density. However, for simplicity, we'll treat the prior as a discrete distribution throughout this paper; small notational changes are required for the prior to be a density.

Now, suppose we have a maximum acceptable probability of mislabeling the input $v_1$; let this parameter be denoted $\beta > 0$. Also, let $u$ denote the true input. The optimal estimator with this constraint is:
\begin{equation}
\label{eq:np_prob}
	\begin{array}{rl}
		\min_{\widehat u} 	& P(\widehat u = v_1 \vert u = v_0) 	\\
		\text{subject to }	& P(\widehat u = v_0 \vert u = v_1) \leq \beta
	\end{array}
\end{equation}
By the Neyman-Pearson lemma, the non-Bayesian detection problem in Equation \ref{eq:np_prob} has the following solution:
\begin{equation}
	\widehat {u}_{\text{NB}}(y) =
	\begin{cases}
		v_1 \text{ if } L(y) \geq \lambda 	\\
		v_0 \text{ otherwise}
	\end{cases}
\end{equation}
where $\lambda$ is chosen such that $P(\widehat {u}_{\text{NB}} = v_0 \vert u = v_1) = \beta$.

Throughout the rest of this paper, we will consider the MAP, but these can be extended to the other two cases. The probability of interest is the probability of successful NILM:
\begin{definition}
	For the two-input case, the \emph{probability of successful NILM} for an estimator $\widehat u$ is:
	\begin{equation}
	\label{eq:2input_pr_succ}
	\sum_{i = 0}^1 P(\widehat u(y) = v_i \vert u = v_i) p( u = v_i )
	\end{equation}
\end{definition}
This can be explicitly calculated given the densities and the prior. 
Additionally, any algorithm and discriminator $(I \circ S)$ will perform worse than $\widehat {u}_{\text{MAP}}$, so the MAP estimate provides an upper bound on any algorithm's probability of successful NILM.

\begin{proposition}
Any estimator $\widehat {u}$ will have a probability of successful NILM bounded by:
	\begin{equation}
	\label{eq:2input_pr_succ}
	\sum_{i = 0}^1 P(\uhatMAP(y) = v_i \vert u = v_i) p( u = v_i )
	\end{equation}
\end{proposition}

\subsection{Distinguishing a finite number of scenarios}

This easily extends to distinguishing between a finite number of scenarios. Let $V$ denote a finite set of inputs. Then:
\begin{definition}
	For the $N$-input case, the \emph{probability of successful NILM} for an estimator $\widehat u$ is:
	\begin{equation}
	\sum_{i = 1}^N P(\widehat u(y) = v_i \vert u = v_i) p( u = v_i )
	\end{equation}
\end{definition}
The MAP is given by:
\begin{equation}
\widehat {u}_{\text{MAP}}(y) =
{\arg \max}_{v \in V} P(G(u,\cdot) = y | u = v) p(u = v)
\end{equation}
\begin{proposition}
There is an upper bound to the probability of successful NILM provided by the MAP:
	\begin{equation}
	\sum_{i = 1}^N P(\uhatMAP(y) = v_i \vert u = v_i) p( u = v_i )
	\end{equation}
\end{proposition}

\subsection{Distinguishing two collections of scenarios}

This philosophy of deriving an upper bound extends nicely to whenever we wish to distinguish two collections of scenarios. This corresponds to distinguishing two sets of inputs.

Now, suppose we have two sets of inputs: $V_0$ and $V_1$. We can still define the probability of successful NILM in this context:
\begin{definition}
	For the case where we wish to distinguish two sets of inputs, the \emph{probability of successful NILM} for an estimator $\widehat u$ is:
	\begin{equation}
	\label{eq:2set_pr_succ}
	\sum_{i = 0}^1 P(\widehat u(y) \in V_i \vert u \in V_i) p(u \in V_i)
	\end{equation}
\end{definition}
Depending on the context, this quantity may be calculable. In other cases, it may be possible to find good approximations or upper bounds. We will see this arise in Section \ref{sec:applications}.

\section{Gaussian case}
\label{sec:applications}
In this section, we instantiate our theory on the special case where our model is a deterministic function with additive Gaussian noise.

\subsection{Two scenarios}
\label{sec:two_inputs}

Suppose our system takes the following form:
\begin{equation}
G(u,\omega) = h(u) + w(\omega)
\end{equation}
where $h : \Uc \rightarrow \R^T$ is a deterministic function and $w$ is a random variable. 
Furthermore, fix any two inputs $v_0, v_1$ which we wish to distinguish, and suppose that $w$ is a zero-mean Gaussian random variable with covariance $\Sigma$. Furthermore, suppose our prior is $p(u = v_0) = p(u = v_1) = 0.5$.

This can encapsulate the case where the uncertainty arises from measurement noise and model error. Referring to our motivating example, suppose that the only difference between $v_0$ and $v_1$ is the presence of a toaster turning on once in $v_1$. The question we are asking is: can we detect the toaster turning on? 

Then, let $f_0$ denote the Normal pdf with mean $h(v_0)$ and covariance $\Sigma$, and similarly let $f_1$ be the Normal pdf with mean $h(v_1)$ and the same covariance $\Sigma$. For shorthand, let $\mu_0 = h(v_0)$ and $\mu_1 = h(v_1)$. 

Since the covariance matrix $\Sigma$ is the same for both random variables, $\uhatMAP$ is determined by a hyperplane. Let $a^\T = \left( \mu_0 - \mu_1 \right)^\T \Sigma^{-1}$ and $b = \frac{1}{2} \left(\mu_1^\T \Sigma^{-1} \mu_1 - \mu_0^\T \Sigma^{-1} \mu_0 \right)$. Then:
\begin{equation}
\uhatMAP(y) =
\begin{cases}
v_1 & \text{ if } a^\T y + b \leq 0 	\\
v_0 & \text{ otherwise}
\end{cases}
\end{equation}
 
 Now, suppose the input is actually $v_0$. That is, $y$ is distributed according to $f_0$.  Then, the signed distance from $y$ to the boundary of the hyperplane is given by $\frac{1}{\|a\|_2}(a^\T y + b)$. This is a linear function of Gaussian random variable, and is thus also a Gaussian random variable. Furthermore, the mean of this random variable will be $\frac{1}{\|a\|_2}(a^\T \mu_0 + b)$, and the variance will be:
 \begin{equation}
 \label{eq:sig_collect}
 \sigma^2 = \frac{1}{\|a\|_2^2} a^\T \Sigma a = \frac{(\mu_0 - \mu_1)^\T \Sigma^{-1} (\mu_0 - \mu_1)}{(\mu_0 - \mu_1)^\T \Sigma^{-2} (\mu_0 - \mu_1)}
 \end{equation}
 Thus, given that the input is actually $v_0$, the probability that $\uhatMAP(y) = v_0$ is:
 \begin{align}
P(\uhatMAP(y) & = v_0 \vert u = v_0) \nonumber \\
 \label{eq:norm_cdf1}
 & = \quad \frac{1}{2} \left( 1 - \operatorname{erf}\left(\frac{-\frac{1}{\|a\|_2}(a^\T \mu_0 + b)}{\sqrt{2\sigma^2}}\right) \right)
\end{align}
 where $\operatorname{erf}$ is the Gauss error function and Equation \ref{eq:norm_cdf1} is simply the 1 minus the cumulative distribution function (cdf) of the distance to the hyperplane evaluated at 0, \ie  the probability that the signed distance is positive.
 
 The computations are exactly the same for the case where the input is $v_1$. 
 Thus:
 \begin{proposition}
  By Equation \ref{eq:2input_pr_succ}, 
 the probability of successfully distinguishing $v_0$ and $v_1$ with the MAP is given by:
 \begin{equation}
 \label{eq:norm_cdf}
\frac{1}{2} \left( 1 - \operatorname{erf}\left(\frac{-\frac{1}{\|a\|_2}(a^\T \mu_0 + b)}{\sqrt{2\sigma^2}}\right) \right)
 \end{equation}
 \end{proposition}
 
 Note that, in general, disaggregation algorithms would not be designed simply to distinguish between $v_0$ and $v_1$, and are likely not to be optimal in this regard. That is, Equation \ref{eq:norm_cdf} provides a theoretical upper bound on how good any possible disaggregation algorithm could perform in distinguishing $v_0$ and $v_1$. Also, note that $\frac{1}{\|a\|_2}(a^\T \mu_0 + b)$ will be positive if $\mu_0 \neq \mu_1$. It follows that the upper bound is always greater than 0.5 if $\mu_0 \neq \mu_1$, and the MAP achieves this upper bound. Thus, if the inputs cause different outputs from the system, there will always exist an algorithm that improves the discrimination between $v_0$ and $v_1$ over blind guessing.
 
 \subsection{N scenarios}
 \label{sec:N_in}

In this subsection, we build on the development in Section~\ref{sec:two_inputs} to handle the case where we wish to distinguish several inputs.

Suppose now that we have a finite set of inputs that we wish to distinguish. Consider the set $\{u_i\}_{i = 1}^{N}$, where $u_i \in \Uc$ for each $i$. Again, suppose all these inputs are equally likely, \ie  $p(u = v_i) = \frac{1}{N}$ for all $i$. We wish to find the MAP. 
We carry over the assumption of Gaussian noise with variance $\Sigma$. The MAP will partition $\R^T$ with hyperplanes of the form given in Section \ref{sec:two_inputs}.

So, suppose the actual input is $u_1$. We wish to ask: what is the probability the MAP will accurately identify $u_1$ from the other $N-1$ inputs? Let $\mu_i = h(u_i)$ for $i = 1,\dots,N$. Then, let $a_i^T = (\mu_1 - \mu_i)^T \Sigma^{-1}$ and $b_i = \frac{1}{2} (\mu_i^T \Sigma^{-1} \mu_i - \mu_1 \Sigma^{-1} \mu_1)$. Given our observation $y \in \R^T$, we wish to ask the probability that $\frac{1}{\|a_i\|_2} (a_i^T y + b_i) > 0$ for $i = 2,\dots,N$, \ie  that the input $u_1$ is more likely than any of the other inputs. More succinctly, define:
\begin{equation}
\label{eq:N_in_mat}
A =
\begin{bmatrix}
a_2^T/\|a_2\|_2 \\
a_3^T/\|a_3\|_2 \\
\vdots \\
a_N^T/\|a_N\|_2
\end{bmatrix} \quad
b = 
\begin{bmatrix}
b_2/\|a_2\|_2 \\
b_3/\|a_3\|_2 \\
\vdots \\
b_N/\|a_N\|_2
\end{bmatrix}
\end{equation}
We wish to ask the probability that $Ay + b$ is in the positive orthant of $\R^N$. Recall that $y$ is distributed according to mean $\mu_1$ and covariance $\Sigma$. Thus, the random variable $Ay + b$ has mean $A\mu_1 + b$ with covariance $A \Sigma A^T$. The probability that this random variable is in the positive orthant cannot be analytically calculated, but can be approximated with high accuracy.


This can be done for $i = 2,\dots,N$ as well, and provide an upper bound on the probability of successful NILM. An example based on real data will be explicated in Section \ref{sec:real_data}.

\subsection{Linear systems}
\label{sec:gauss_lin}
In this subsection, we specialize the previous theory to the case where all our devices are linear systems. Suppose that the dynamics of our household are of the form $y = Au + e$, and our noise $e$ has covariance $\widehat \sigma^2 I$. Note that $\sigma^2$ as defined in Equation~\ref{eq:sig_collect} is equal to $\widehat \sigma^2$.

Now, suppose the sets that we wish to distinguish are $V_0 = \{0\}$ and $V_1 = \{ v : L \leq \|v\|_2 \leq U \}$, for some constants $0 < L \leq U$. That is, can we detect an input with magnitude in the range $[L,U]$? By Equation \ref{eq:2set_pr_succ}, we have the probability of successful NILM for an estimator $\widehat u$ is:
\begin{equation}
P(\widehat u(y) = 0 \vert u = 0 )p(u = 0) + 
P(\widehat u(y) \in V_1 \vert u \in V_1)p(u \in V_1)
\end{equation}

First, consider a fixed input $v \in V_1$. If we suppose that $u  = v$, then the probability of an estimator $\widehat u$ distinguishing $v$ from 0 is bounded by:
\begin{equation}
P(\widehat u(y) \neq 0 \vert u = v) \leq \frac{1}{2} \left( 1 + \operatorname{erf}\left(\frac{\|Av\|_2}{2\sqrt{2\sigma^2}}\right) \right)
\end{equation}
This can be seen by noting that, after a projection into one-dimension, the separating hyperplane is the point $\pm \|Av\|_2/2$. Without loss of generality, let us suppose the separating point is $\|Av\|_2/2$.

Note that this equation is an increasing function of $\|Av\|_2$. This gives us:
\begin{equation}
\frac{1}{2} \left( 1 + \operatorname{erf}\left(\frac{\|Av\|_2}{2\sqrt{2\sigma^2}}\right) \right) \leq
\frac{1}{2} \left( 1 + \operatorname{erf}\left(\frac{\sigma_{\max}(A) U }{2\sqrt{2\sigma^2}}\right) \right)
\end{equation}
where $\sigma_{\max}(A)$ is the largest singular value of $A$. This held for any $v \in V_1$, so measure-theoretic properties give us:
\begin{equation}
P(\widehat u(y) \in V_1 \vert u \in V_1) \leq 
\frac{1}{2} \left( 1 + \operatorname{erf}\left(\frac{\sigma_{\max}(A) U }{2\sqrt{2\sigma^2}}\right) \right)
\end{equation}

\begin{proposition}
In the linear system case, the probability of successful NILM is bounded above by:
\begin{equation}
p(u = 0) + 
\frac{1}{2} \left( 1 + \operatorname{erf}\left(\frac{\sigma_{\max}(A) U }{2\sqrt{2\sigma^2}}\right) \right)
p(u \in V_1)
\end{equation}
\end{proposition}
Thus, even with just knowledge of the variance of the noise and the sensitivity of our linear systems, we can still find an upper bound on the probability of successful NILM.

\section{Real data analysis}
\label{sec:real_data}
In this section, we take the theory from Sections \ref{sec:theory} and \ref{sec:applications} and use them on real data to address several different problems. We used the emonTx wireless open-source energy monitoring node from OpenEnergyMonitor\footnote{ {\tt{http://openenergymonitor.org/emon/emontx}}} to take RMS current measurements at 12Hz. 

\begin{figure}[!ht]
	\begin{center}
	\includegraphics[width=\columnwidth]{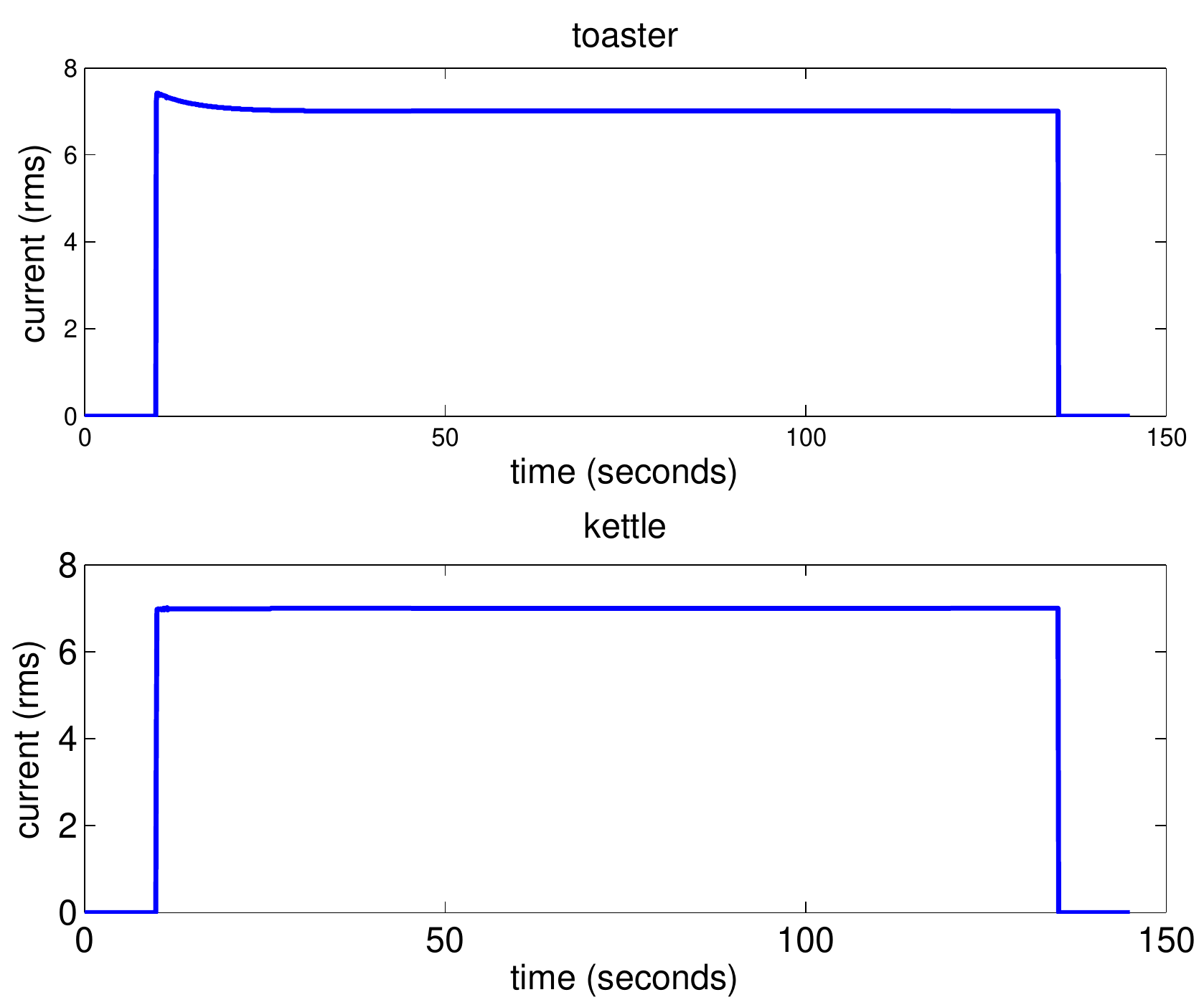}
	\end{center}
	\caption{\emph{Top: }The measured RMS current signal for a toaster. \emph{Bottom: }The measured RMS current signal for a kettle.}
	\label{fig:toaster_sig}
\end{figure}

\begin{figure}[!ht]
	\begin{center}
	\includegraphics[width=\columnwidth]{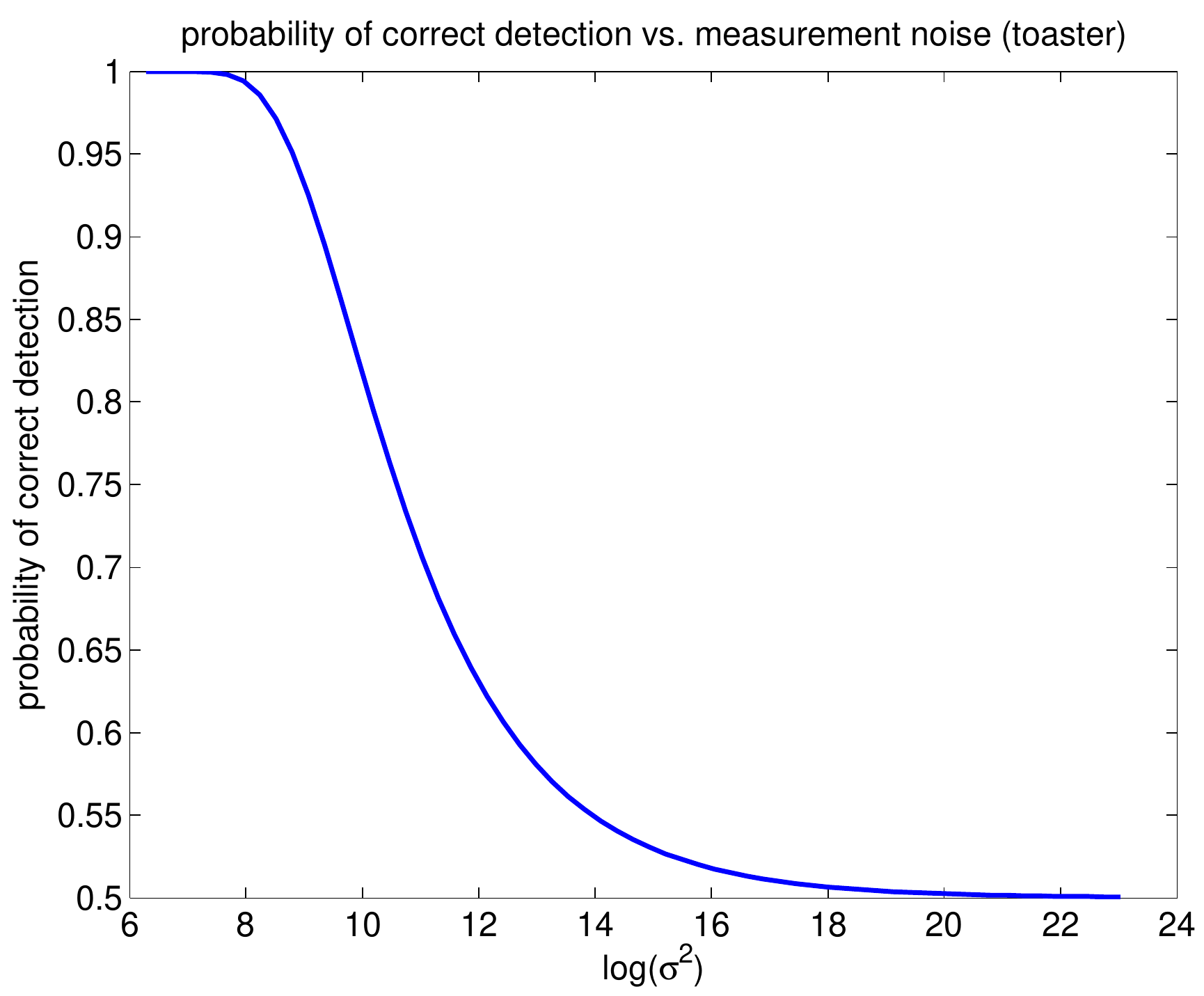}
	\end{center}
	\caption{The probability of successful identification of a toaster as a function of modeling and measurement error.}
	\label{fig:prob_toaster}
\end{figure}

\begin{problem}
What is an upper bound for the probability of successfully detecting a toaster turning on, as a function of the modeling and measurement error?
\end{problem}
We took measurements from a toaster. The basic signal is shown in Figure \ref{fig:toaster_sig}.
We use the assumptions outlined in Section \ref{sec:two_inputs}. Additionally, we let our emonTx measurements serve as ground truth, and assumed that the covariance of the Gaussian noise was $\sigma^2 I$, \ie  the noise was uncorrelated at each time step. Following the analysis in Section \ref{sec:two_inputs}, the probability of distinguishing the toaster turning on is shown in Figure \ref{fig:prob_toaster}.

Note that $\sigma^2$ has to grow considerably large before the optimal algorithm starts to fail to distinguish the toaster from nothingness. This is unsurprising, as the optimal algorithm would have several samples to distinguish quite separate means.

\begin{figure}[!ht]
	\begin{center}
	\includegraphics[width=\columnwidth]{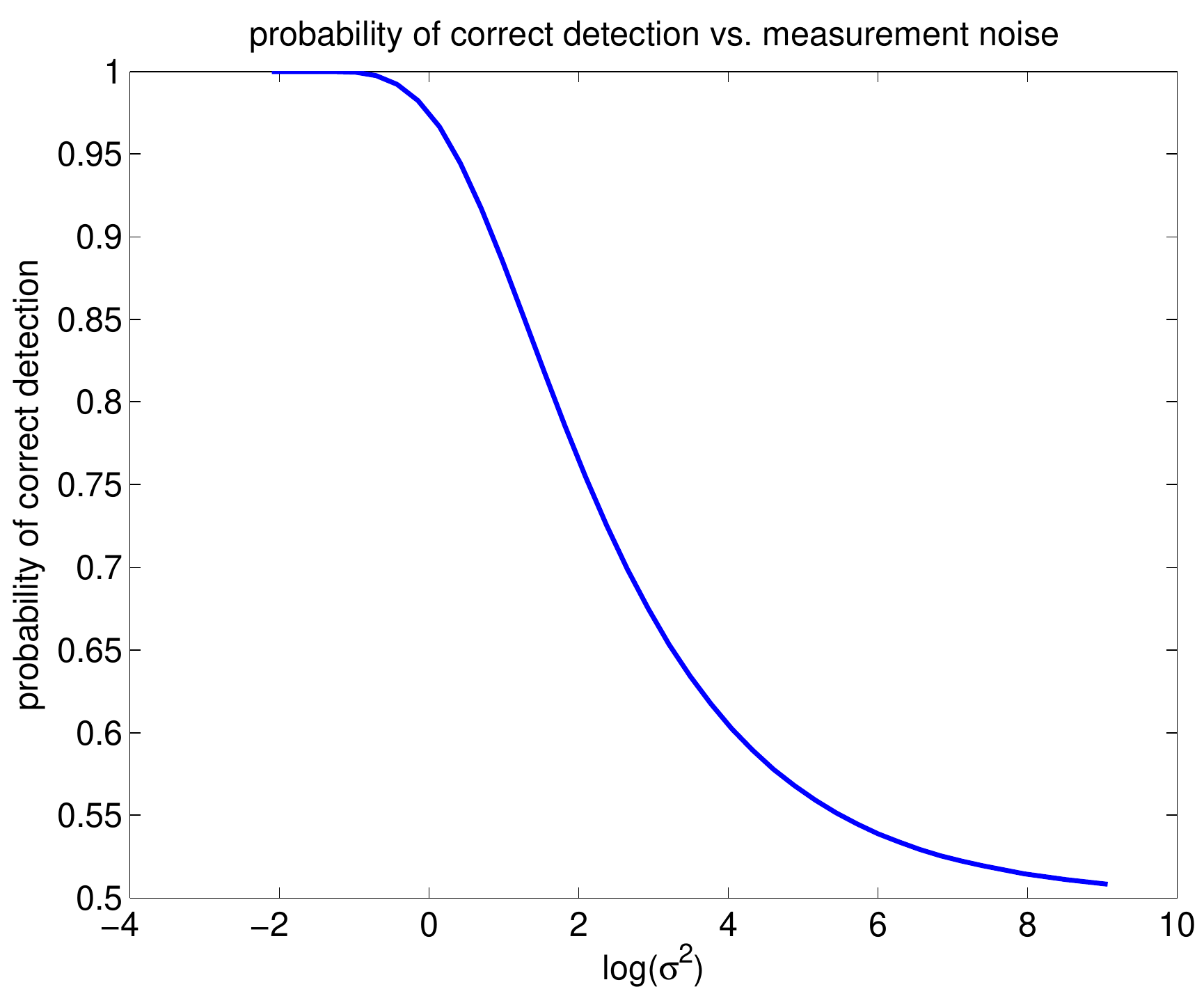}
	\end{center}
	\caption{The probability of successfully discriminating of a toaster and a kettle as a function of modeling and measurement error.}
	\label{fig:t_v_k}
\end{figure}

\begin{problem}
\label{prob:tvk}
What is an upper bound for the probability of successfully distinguishing a toaster turning on and a kettle turning on, as a function of the modeling and measurement error?
\end{problem}
We repeated this analysis with both a toaster and a kettle signal, depicted in Figure \ref{fig:toaster_sig}. The devices are on for exactly the same time window. The results are shown in Figure~\ref{fig:t_v_k}. As we can see, the variance on the error is orders of magnitude smaller when the probability drops to near 0.5. However, the $\sigma^2$ value is still quite large, and we likely can distinguish the two devices at 12Hz.

\begin{figure}[!ht]
	\begin{center}
	\includegraphics[width=\columnwidth]{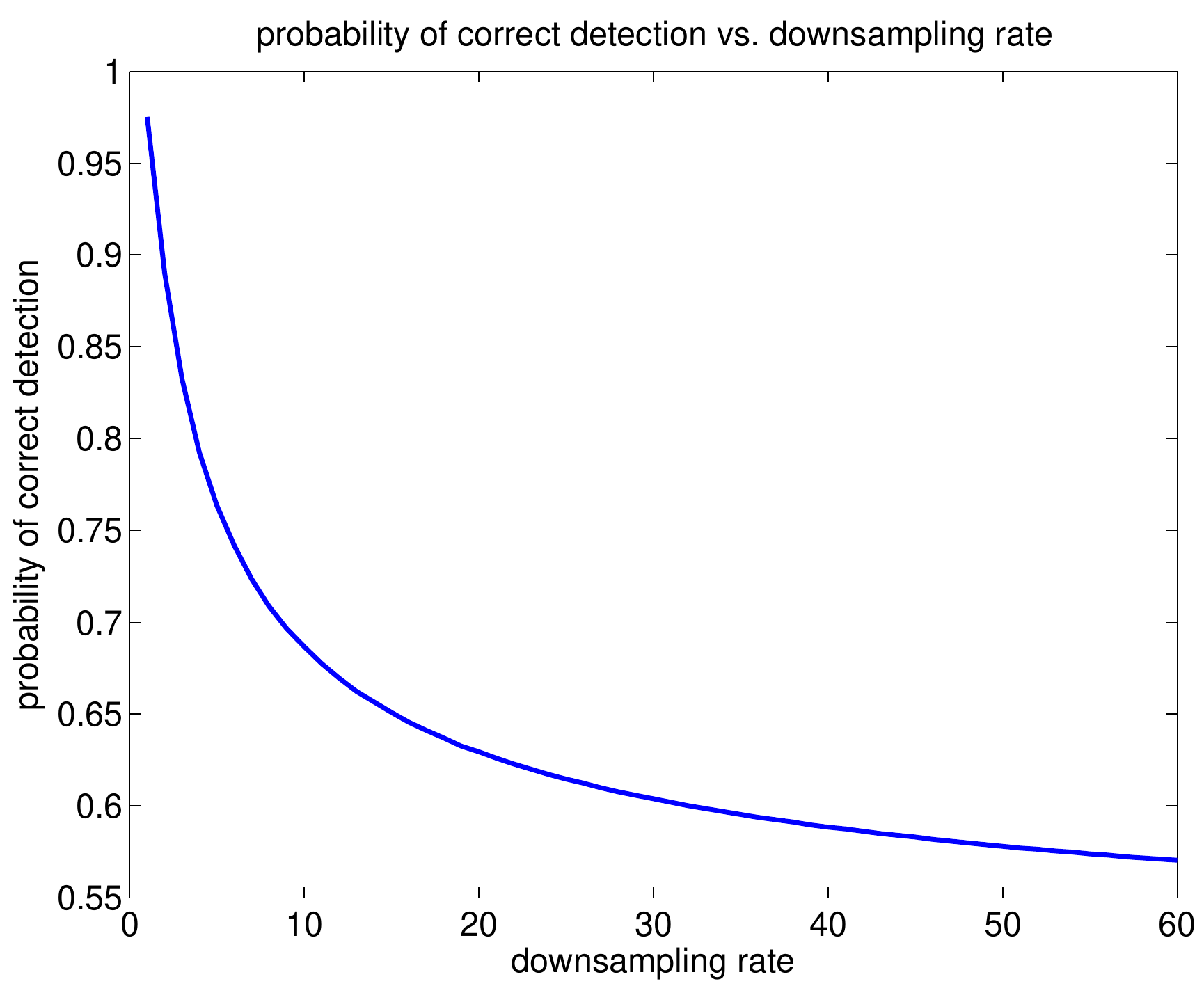}
	\end{center}
	\caption{The probability of successfully discriminating a toaster and a kettle as a function of the sampling rate. We fixed $\sigma^2 = 1$.}
	\label{fig:tvk_samp}
\end{figure}

\begin{problem}
What is an upper bound for the probability of successfully distinguishing a toaster turning on and a kettle turning on, as a function of the sampling rate?
\end{problem}
The results to Problem \ref{prob:tvk} are promising, as they tell us it is very possible to distinguish two rather similar devices. However, the sampling rate of 12Hz is very high. Now, we analyze how likely we are to distinguish the two devices as the sampling rate changes. This is shown in Figure \ref{fig:tvk_samp}. We down-sampled the 12Hz signal. Additionally, if we down-sampled with rate $K$, we assumed it was equally likely that the signal would begin on any of the first $K$ time-steps.

As expected, the probability of successful NILM decreases with the sampling rate. Additionally, the performance degrades quite quickly, and we barely perform better than guessing when the downsampling rate is 60, i.e. we sample every 5 seconds. 
This result allows us to determine a lower bound on the sampling rate necessary to achieve a certain effectiveness of NILM. It gives prescriptions on what hardware specifications and network capacity is needed in AMIs to achieve a certain goal.

\begin{problem}
What is an upper bound for the probability of successfully distinguishing several devices, as a function of measurement and modeling error?
\end{problem}

\begin{figure}[!ht]
	\begin{center}
	\includegraphics[width=\columnwidth]{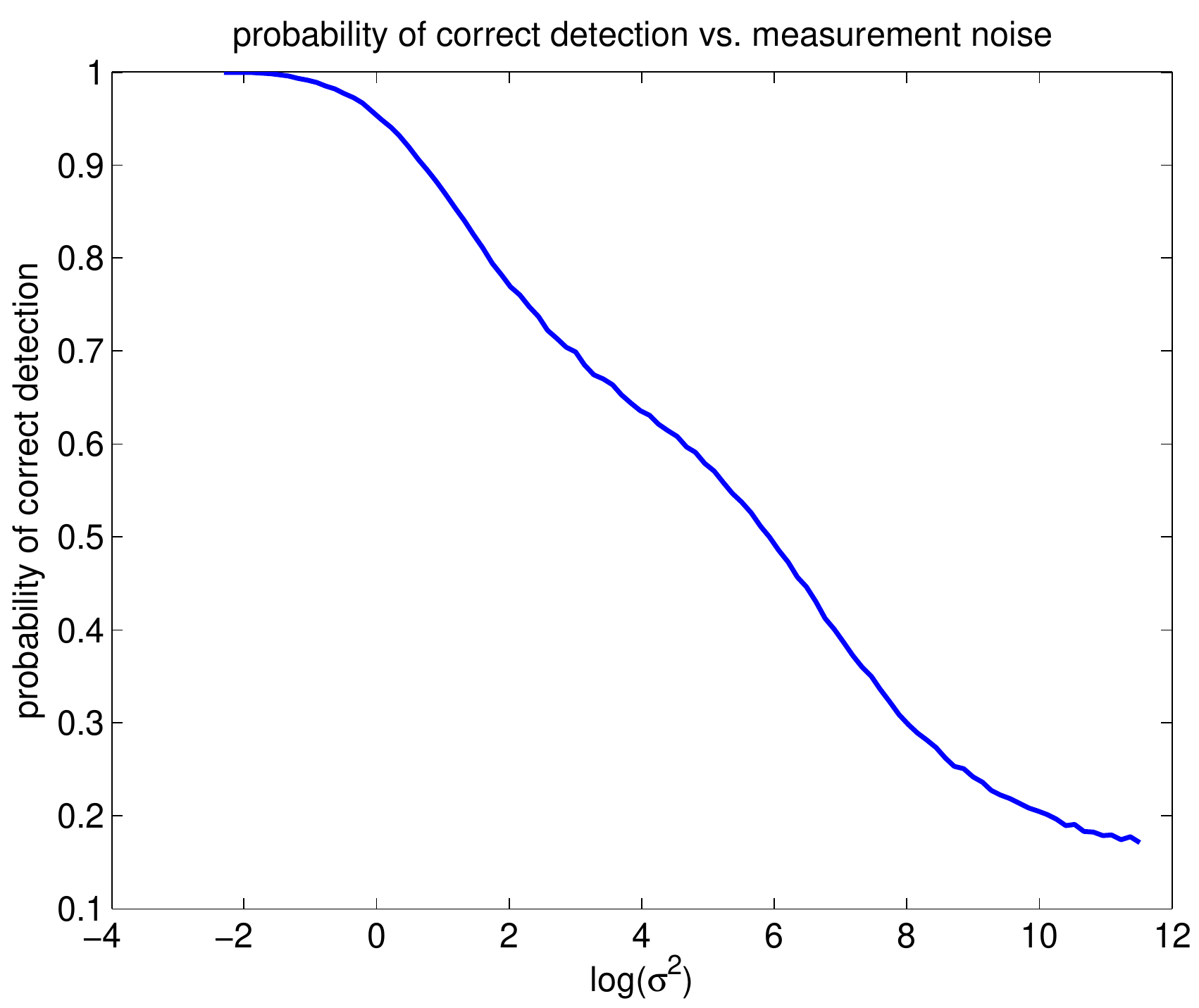}
	\end{center}
	\caption{The probability of successfully discriminating one device from the $N-1$ other devices as a function of the measurement and modeling error.}
	\label{fig:N_dev}
\end{figure}

Here, we use the results in Section \ref{sec:N_in}. The devices in question are a microwave, a toaster, a kettle, an LCD computer monitor, a projector, and a digital oscilloscope. As before, we have one signal for each device, which is activated for the same time window for each device. That is, for each device, we have a \emph{fixed} input. If each device is equally likely to turn on, we have the results shown in Figure \ref{fig:N_dev}.

\begin{problem}
What is an upper bound for the probability of successfully distinguishing several linear systems, as a function of the input magnitude?
\end{problem}
Suppose we have the same 6 devices as in the previous problem. Furthermore, suppose they are linear systems, and the observed signals were a result of an input which was a pulse of magnitude 1. Then, we can use results from Section \ref{sec:gauss_lin}. 

\begin{figure}[!ht]
	\begin{center}
	\includegraphics[width=\columnwidth]{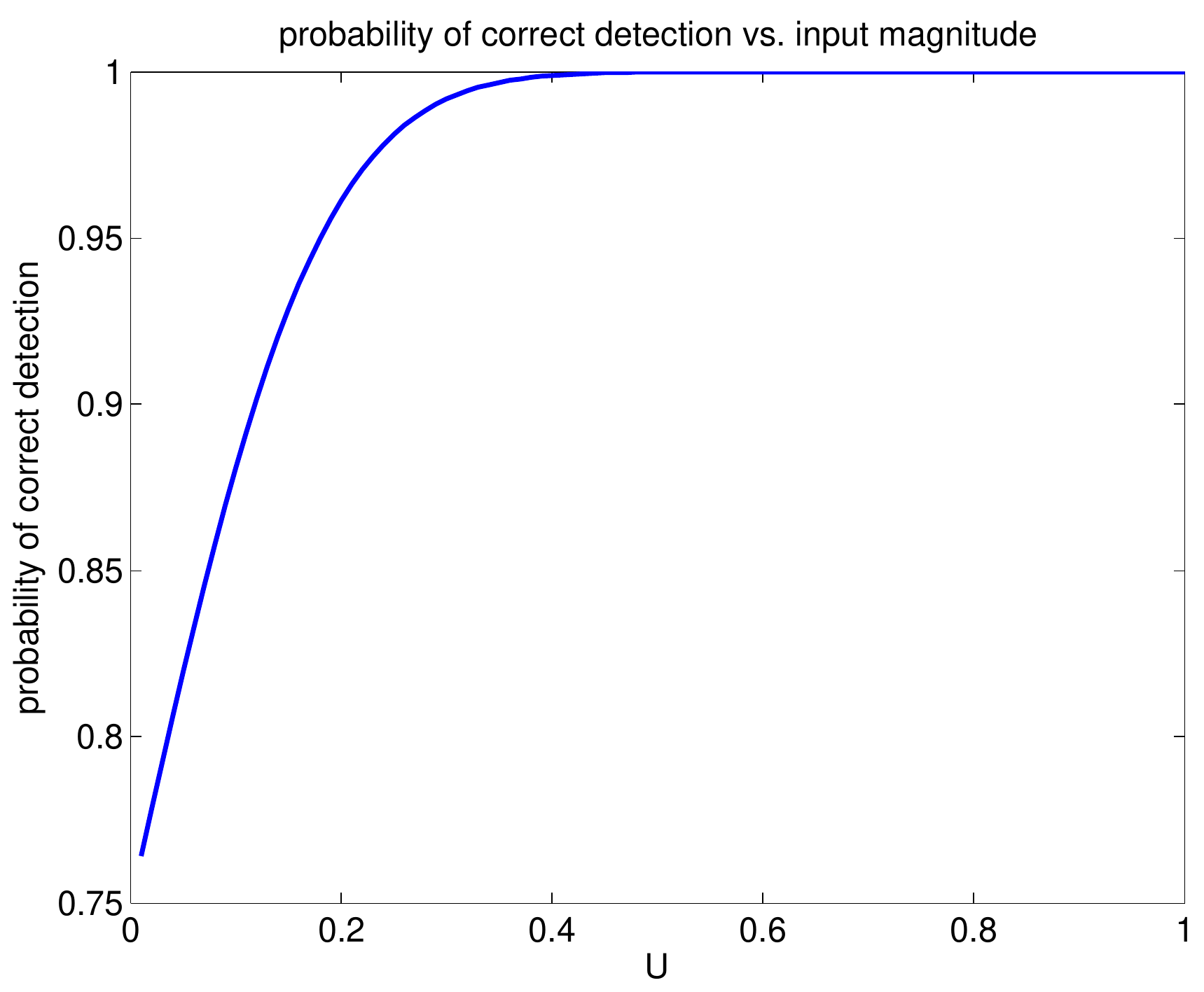}
	\end{center}
	\caption{The probability of successfully discriminating a device turning on from the null hypothesis as a function of the input magnitude $U$. We fix $\sigma^2 = 1$.}
	\label{fig:lin}
\end{figure}

Suppose we wish to determine whether or not a device turned on. The input to our device is nonzero and bounded by $U$. We plot the upper bound from Section \ref{sec:gauss_lin} on the probability of successful NILM as a function of $U$. The results are in Figure \ref{fig:lin}.


\section{Conclusion}
\label{sec:conclusion}
In this paper, we explore the fundamental limits of NILM algorithms. More specifically, we derive an upper bound on the probability of distinguishing scenarios for an arbitrary NILM algorithm. First, we present the theory in its general case, and then we instantiate the theory on the case where the error is additive Gaussian noise independent of the underlying scenario. With this upper bound in hand, and our Gaussian assumption, we interpret real data we collected and discuss how the probability of successful NILM depends on the modeling and measurement error, the sampling rate, and the magnitude of the device usage.

To the best of our knowledge, this is the first paper investigating the fundamental limits of NILM. These fundamental limits are useful for several reasons. They can provide a guarantee on when NILM is impossible, which has implications for the design of privacy-aware AMIs, as well as privacy policies in AMIs. These limits are algorithm-independent, so they will hold regardless of changing technologies. These limits also can provide prescriptions for the design of AMIs, if NILM of a certain sort is desired, in terms of network capacity and sensor accuracy. Finally, it also provides a unified framework for understanding the problem of NILM.

\section{Acknowledgments}
The authors would like to thank Alvaro C\'ardenas for his helpful comments on an early draft of this document, as well as Aaron Bestick for stimulating conversations and assistance with our experimental setup.

%
\bibliographystyle{abbrv}
\bibliography{DONG_ROY_refs}  
%

\end{document}